\documentstyle[11pt,paspconf,psfig]{article}

\def\apj{ApJ }

\def\aa{A\&A }

\def\pasp{PASP }

\def\iau143{1991, in ``Wolf-Rayet stars and Interrelations with Other Massive Stars in 
	Galaxies'', IAU Symp.~143, Eds.~K.A.~Van der Hucht, B.~Hidayat, Kluwer}
\def\iau145{YEAR, in ``Evolution of Stars: the photospheric abundance 
	connection'', IAU Symp.~145, Eds.~G.~Michaud, A.~Tutukov, Kluwer}
\def\iau163{1994, in ``Wolf-Rayet stars: Colliding winds, binaries, evolution'',
	Eds.~K.A.~van der Hucht, P.M.~Williams, Kluver Academic Publ., Dordrecht}

\def\etal{et al.~}

\def\pder#1#2{\ifmmode  {\frac{\partial #1}{\partial #2}}\fi}
\def\lder#1{\ifmmode  {\frac{D #1}{Dt}}\fi}

\def\sun{\hbox{$\odot$}} 
\def\la{\mathrel{\mathchoice {\vcenter{\offinterlineskip\halign{\hfil
$\displaystyle##$\hfil\cr<\cr\noalign{\vskip1.5pt}\sim\cr}}}
{\vcenter{\offinterlineskip\halign{\hfil$\textstyle##$\hfil\cr<\cr
\noalign{\vskip1.0pt}\sim\cr}}}
{\vcenter{\offinterlineskip\halign{\hfil$\scriptstyle##$\hfil\cr<\cr
\noalign{\vskip0.5pt}\sim\cr}}}
{\vcenter{\offinterlineskip\halign{\hfil$\scriptscriptstyle##$\hfil
\cr<\cr\noalign{\vskip0.5pt}\sim\cr}}}}}
\def\ga{\mathrel{\mathchoice {\vcenter{\offinterlineskip\halign{\hfil
$\displaystyle##$\hfil\cr>\cr\noalign{\vskip1.5pt}\sim\cr}}}
{\vcenter{\offinterlineskip\halign{\hfil$\textstyle##$\hfil\cr>\cr
\noalign{\vskip1.0pt}\sim\cr}}}
{\vcenter{\offinterlineskip\halign{\hfil$\scriptstyle##$\hfil\cr>\cr
\noalign{\vskip0.5pt}\sim\cr}}}
{\vcenter{\offinterlineskip\halign{\hfil$\scriptscriptstyle##$\hfil
\cr>\cr\noalign{\vskip0.5pt}\sim\cr}}}}}
\def\masl{\ifmmode  {\rm M_{\sun}yr^{-1}} \else ${\rm M_{\sun}yr^{-1}}$\fi}

\def\mdot{\ifmmode  \dot{M} \else $\dot{M}$\fi}
\def\msun{\ifmmode M_{\odot} \else $M_{\odot}$\fi}
\def\vinf{\ifmmode v_{\infty} \else $v_{\infty}$\fi}
\def\teff{\ifmmode T_{\rm eff} \else $T_{\rm eff}$\fi}
\def\logg{\ifmmode \log g \else $\log g$\fi}
\def\loggeff{\ifmmode \log g_{\rm eff} \else $\log g_{\rm eff}$\fi}
\def\rstar{\ifmmode R_{\star} \else $R_{\star}$\fi}
\def\lstar{\ifmmode L_{\star} \else $L_{\star}$\fi}
\def\mstar{\ifmmode M_{\star} \else $M_{\star}$\fi}
\def\rsun{\ifmmode R_{\odot} \else $R_{\odot}$\fi}
\def\lsun{\ifmmode L_{\odot} \else $L_{\odot}$\fi}
\def\12c16o{$^{12}{\rm C}\left(\alpha,\gamma\right)^{16}{\rm O}$}
\def\kms{\ifmmode {\rm km \;s^{-1}} \else $\rm km \;s^{-1}$\fi}
\def\nlte{non--LTE}
\def\lb{line blanketing}

\def\taustar{\ifmmode \tau_{\star} \else $\tau_{\star}$\fi}
\def\tauross{\ifmmode \tau_{\rm Ross} \else $\tau_{\rm Ross}$\fi}
\def\teffstar{\ifmmode{T^{\star}_{\rm eff}} \else $T^{\star}_{\rm eff}$\fi}

\def\ang{\ifmmode {\rm \AA} \else $\rm \AA$\fi}  

\def\hii{H~{\sc ii}}
\def\hei{He~{\sc i}}
\def\heii{He~{\sc ii}}

\def\he0{\ifmmode {\rm He^{\circ}} \else $\rm He^{\circ}$\fi}
\def\hep{\ifmmode {\rm He^+} \else $\rm He^{+}$\fi}
\def\hepp{\ifmmode {\rm He^{2+}} \else $\rm He^{2+}$\fi}
\def\halpha{\ifmmode {\rm H{\alpha}} \else $\rm H{\alpha}$\fi}
\def\hgamma{\ifmmode {\rm H{\gamma}} \else $\rm H{\gamma}$\fi}
\input rotate
\def\rotfig#1#2#3{\epsfxsize=#3cm \newbox\rotbox \setbox\rotbox=\hbox{\epsfbox{#2}}\centering\noindent\rotr\rotbox    \vspace{-0.4cm}}
\begin{document}

\title{The ionizing fluxes of early type stars and their impact on \hii\ regions 
and the `Diffuse Ionized Gas'} 
\author{Daniel Schaerer}
\affil{Space Telescope Science Institute, 3700 San Martin Drive, Baltimore, 
	MD 21218, USA (e--mail: schaerer@stsci.edu)}

\begin{abstract}
We discuss recent results on the ionizing fluxes of O stars obtained from
our ``combined stellar structure and atmosphere models'' ({\em CoStar}) 
accounting for stellar winds, non-LTE effects and \lb.
The implications on the ionization structure of \hii\ regions are summarized,
and observational constraints on the ionizing spectra and the total ionizing
photon fluxes are presented.
Using our {\em CoStar} models we derive new consistent predictions for the
H and \hei\ ionizing fluxes of steady-state massive star populations.
Implications for the interpretation of observations of `Diffuse Ionized
Gas' in galaxies are discussed. 
Finally we present preliminary model calculations aiming to improve our 
current predictions further and compare our results to recent results
from the Munich group.
\end{abstract}


\section{Introduction}
The ``combined stellar structure and atmosphere models'' 
(hereafter {\em CoStar}) models have been applied in detail 
to the main-sequence phase (Schaerer \etal 1996ab; paper I \& II).
Here we concentrate on some important predictions regarding the spectral evolution
on the main sequence, namely recent results on the ionizing fluxes of OB stars
for which considerable progress has been achieved in the recent years.
Predictions of ionizing fluxes are of special interest for analysis of massive
stars, \hii\ regions, young starburst galaxies and related objects.

We shall first present current models for O stars (Sect.\ \ref{s_models}).
In Sect.\ \ref{s_hii} we will summarize some of the implications recent
ionizing fluxes have on \hii\ regions and how observations of \hii\ regions
might be used to constrain model atmospheres.
New predictions of ionizing fluxes of steady-state massive star populations
and some of their implications on the diffuse ionized gas in galaxies
are given in Sect.\ \ref{s_pops}
In Sect.\ \ref{s_new} we finally present preliminary results from new 
calculations aiming to improve our current predictions further.

\section{Ionizing fluxes O stars}
\label{s_models}
The input physics of the {\em CoStar} models is described in detail in 
paper I. 
The most important ingredients regarding the prediction of ionizing
fluxes are the treatment of \nlte, the stellar wind, and line blanketing
(see Schaerer \& de Koter 1997, hereafter paper III).
The current sets of models (paper III) covers the entire main 
sequence evolution
for stars with initial masses between 20 and 120 \msun\
and includes a set for solar metallicity (Z=0.02) and low metallicity
(Z=0.004 and appropriately scaled wind parameters).
The coverage approximately corresponds to spectral types from O3 to B0
and all luminosity classes. 
As discussed in paper III, reliable predictions for later types 
are currently not available for physical and technical reasons.
Spectral energy distributions, \hii\ region models (cf.\ below), and 
additional information is available on our Web page at
{\tt http://www.stsci.edu/ftp/science/\-starburst} and on CD-ROM
(see Leitherer \etal\ 1996).

Extensive comparisons of our predictions with previous calculations
(Kurucz models, plane parallel line blanketed models) are given
in paper III. A detailed discussion of the physical processes 
affecting the ionizing spectrum is also given there.
The main results can be summarized as follows:  
{\em 1)} The flux in the \heii\ continuum is increased 
	by 2 to 3 (3 to 6) orders of magnitudes compared to predictions from 
	plane parallel \nlte\ (LTE) model atmospheres (cf.\ Gabler \etal 1989).
{\em 2)} The flux in the \hei\ continuum is known to be increased due to
	\nlte\ effects. In addition, the combined effect of a mass outflow 
	and \lb\ leads to a flatter energy distribution in the \hei\ continuum.
	This has profound implications on the structure of
	\hii\ regions (Sect.\ \ref{s_hii}).
{\em 3)} To a lesser extent than at high energies, the flux in the Lyman 
	continuum is also modified due to \lb\ and the presence of a stellar
	winds. Generally one obtains a reduction of the Lyman continuum 
	flux compared to plane parallel model atmospheres.

Based on our {\em CoStar} models we have also derived integrated
photon fluxes for stellar parameters corresponding to the
new spectral type and luminosity class calibration of Vacca \etal\
(1996). 
Our results are compared with former calibrations of Panagia (1973) and
Vacca \etal (1996), both based on plane parallel model atmospheres.
It must, however, be remembered that such a calibration is quite strongly
dependent on the assignment between spectral types and stellar parameters,
for which considerable uncertainties remain for O stars 
(cf.\ Vacca et al., Crowther 1997, Lamers et al.\ 1997, Schmutz, 
these proceedings).

\section{The impact of {\em CoStar} ionizing fluxes on \hii\ regions
and possible tests of atmosphere models }
\label{s_hii}
Using the photoionization code PHOTO 
we have computed a large grid of
single star \hii\ region models based on the new {\em CoStar} ionizing 
fluxes (Stasi\'nska \& Schaerer 1997, hereafter paper IV). For comparisons a 
similar grid has been calculated using fluxes from plane parallel LTE 
model atmospheres of Kurucz (1991). 
The main additional parameter determining the structure of the \hii\
region, the ionization parameter $U$, is varied over a wide range
corresponding to conditions expected between the most compact \hii\ regions
and the most diluted interstellar medium.
Detailed results are available on our Web page and on CD-ROM (see above).

\begin{figure*}[htb]
\centerline{
\psfig{width=6.5cm,figure=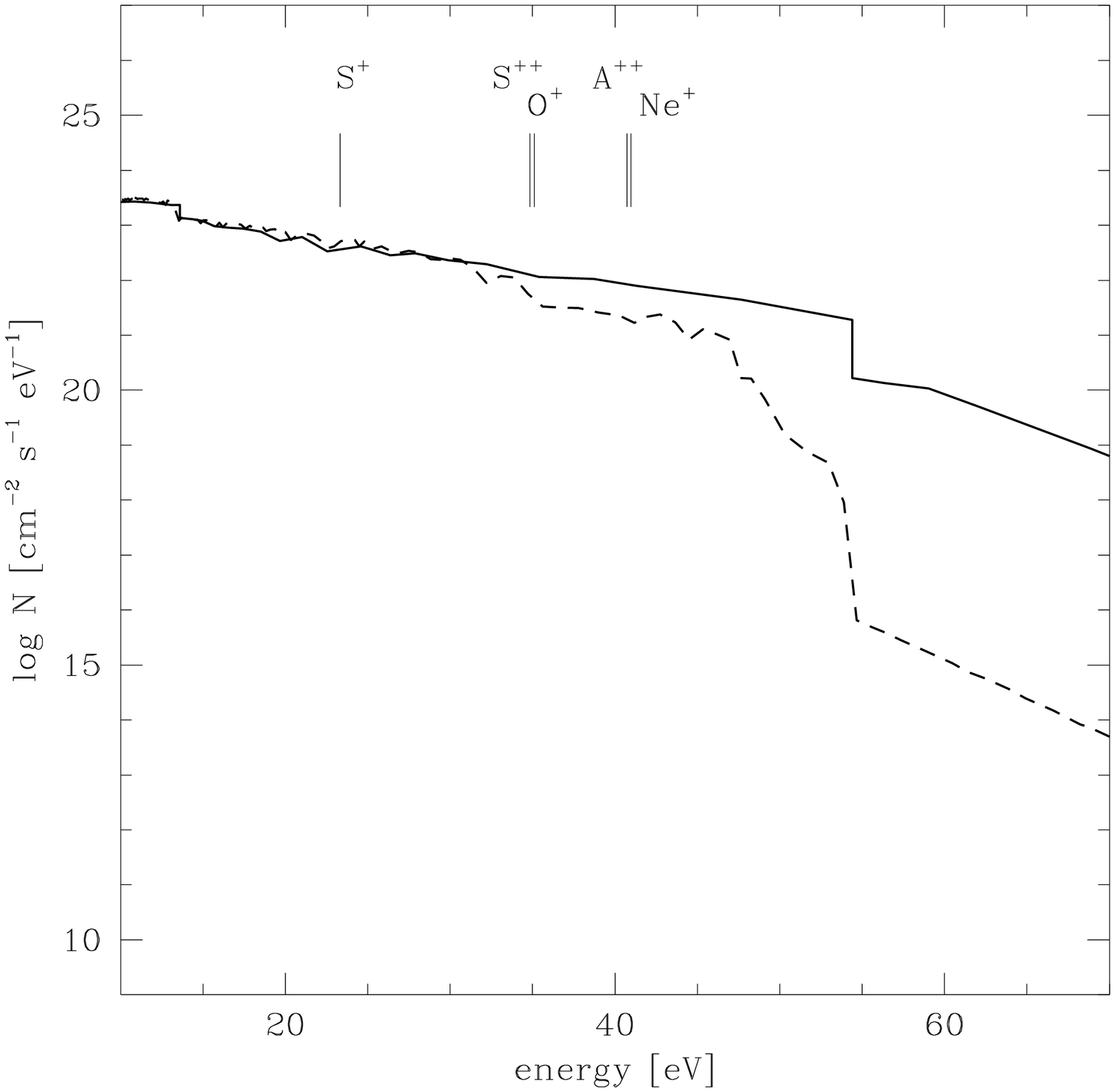}
\psfig{width=6.5cm,figure=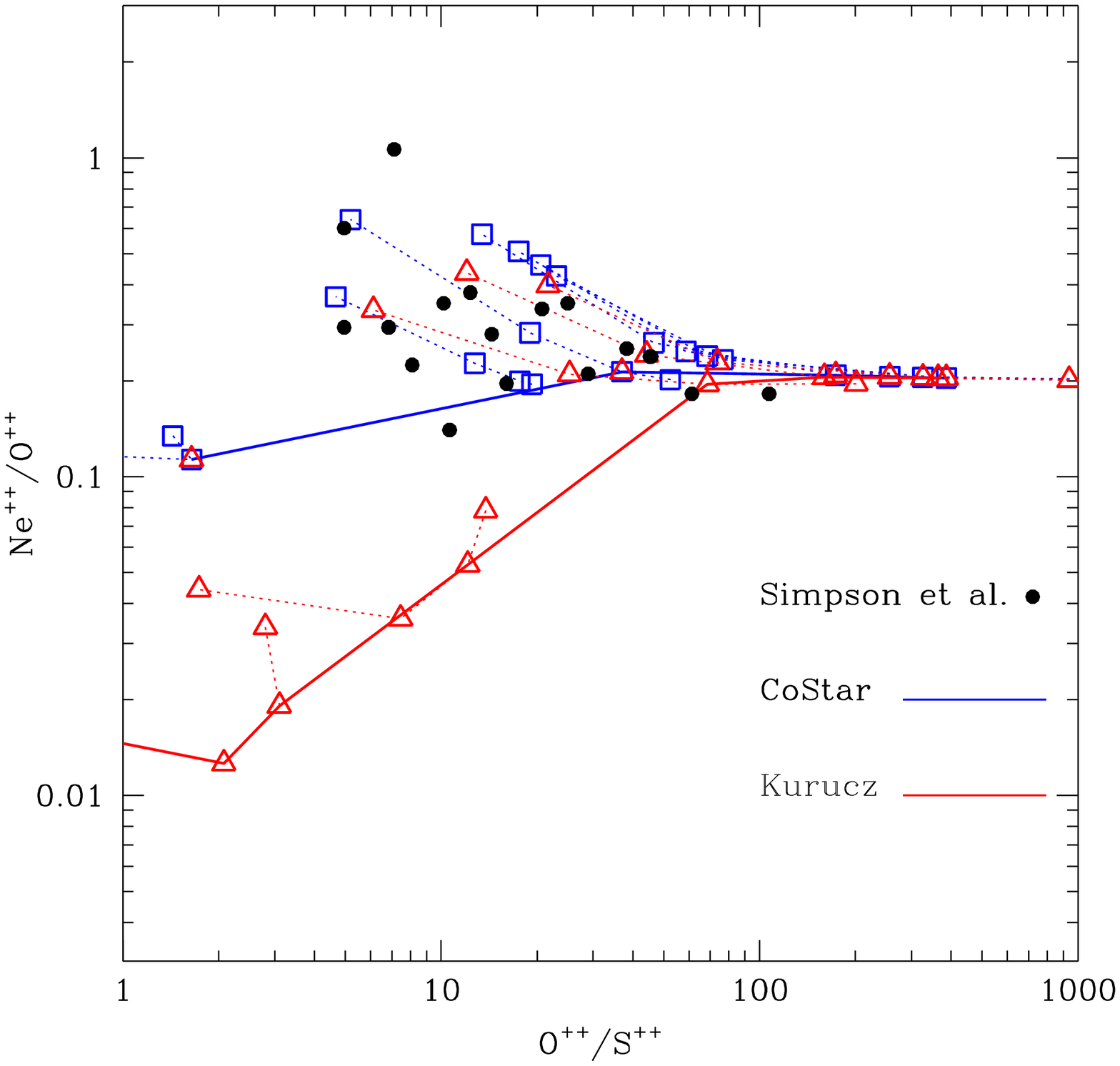}
}
\caption{{\em Left panel:} Comparison of the EUV continuous spectral 
energy distribution from a {\em CoStar} model (solid line) and a 
Kurucz model (dashed) for \teff\ $\sim$ 45000K (from paper IV).
{\em Right panel:}
Predicted ionic Ne$^{++}$/O$^{++}$ abundance ratio as a function
of the O$^{++}$/S$^{++}$ ionic abundance ratio for the solar metallicity 
models (squares: {\em CoStar} models, triangles: Kurucz models).
The solid lines connects photoionization models for a ``standard'' ionization 
parameter (cf.\ paper IV) and varying stellar temperatures.
Dotted lines show variations of the ionization parameter.
Observations from the IR line measurement of Simpson et al.\ (1995) are 
shown by filled circles.}
\label{fig_hii}
\end{figure*}

The flatter spectral energy distribution in the \hei\ continuum 
(cf.\ Fig.\ \ref{fig_hii} left panel)
generally leads to higher ionic ratios in the nebula if {\em CoStar} 
models are compared to plane parallel atmospheres at the same stellar 
temperatures. 
In particular, with ionizing fluxes from {\em CoStar} models one obtains
ionic fractions of $x({\rm N}^+)/x({\rm O}^+)$ and $x({\rm Ne}^{++})/x({\rm O}^{++})$
which are close to unity for ionization parameters typical of \hii\ regions
and over a large range of stellar temperatures. Therefore 
N$^+$/O$^+$ and Ne$^{++}$/O$^{++}$ should provide safe indicators to derive
N/O and Ne/O abundance ratios over a wide range of astrophysical conditions.

As shown by the exploratory calculations of Sellmaier \etal\ (1996) 
the use of ionizing fluxes from \nlte\ wind models including line blocking
seems to resolve the so-called [Ne~{\sc iii}] problem, which was made particularly
clear when Simpson \etal\ (1995) plotted the values of Ne$^{++}$/O$^{++}$ as a 
function of O$^{++}$/S$^{++}$ (all derived from far-IR line measurements). 
The problem is illustrated in Fig.\ \ref{fig_hii} (right panel), where
the observations indicate an essentially flat Ne$^{++}$/O$^{++}$  
while the predictions of photoionizaton models using the Kurucz atmospheres 
fail to reproduce the observations.
Indeed, the higher ionic ratios obtained from photoionization
models using the {\em CoStar} fluxes are in much better agreement with the 
observations than models based on Kurucz atmospheres (see Fig.\ \ref{fig_hii}). 
As shown in paper IV a similar ``observational test'' can be performed by using 
optical lines. Both comparisons support our ionizing fluxes from \nlte\ atmospheres 
including stellar winds and \lb. 

Although such comparisons may represent extremely useful tests of predicted EUV fluxes,
good constraints on the ionization parameter and a precise knowledge about the ionizing
sources (stellar type, number of stars) must be obtained for accurate comparisons.
This should be possible with future observations of, in the ideal case, integrated 
single star \hii\ regions. Careful studies of the metallicity dependence 
(and hence also the dependence on the wind properties) of the predicted fluxes 
will be both necessary and extremely useful for many astrophysical applications.
For example the great sensitivity of the Ne (and several other IR) lines on the 
shape of the
ionizing spectrum in the \hei\ continuum has important consequences for the 
interpretation of emission line spectra of \hii\ regions and young starbursts 
in the wavelength range observed by ISO. 
Derived stellar temperatures, IMF slopes and upper mass cut-offs
depend strongly on the adopted ionizing fluxes as illustrated by 
Kunze et al.\ 1996.

\subsection{Constraints on the Lyman continuum production of OB stars}
\label{s_lyc}
Another fundamental test atmosphere models have to be subjected to is of
course whether the integrated ionizing photon flux (e.g.\ in the Lyman
continuum) is correctly predicted.
The best example of such a test is certainly the B2II star $\epsilon$ CMa,
for which direct EUVE observations below the Lyman limit could be obtained
(Cassinelli et al.\ 1995). For this case a large excess (by a factor of $\sim$ 30)
of the Lyman continuum flux was found compared to plane parallel atmosphere
models ! Although this excess can be explained by a temperature excess in the
continuum forming layers (see Cassinelli et al., paper IV, Aufdenberg et al.\ 
1997) the physical cause for this temperature increase remains to be
explained. 
Obviously in a larger context the fundamental question is, however,
if the Lyman continuum excess is a general phenomenon, and if so 
for what type of stars ?

In principle this question can easily be answered by observations 
measuring the integrated recombination rate (e.g.\ \halpha\ luminosity) 
of nebulae around single stars of known spectral type.
Detection of excess radiation compared to predictions for the determined 
spectral type could indicate an intrinsic excess of the Lyman continuum flux
from the central star (or an undetected additional source).
A lack of radiation can be interpreted as leakage of photons out of the
region or important internal absorption by dust.
To our knowledge no such systematic comparison has been done so far,
especially for B stars. Few statements can nevertheless already be made
at this stage.

For ultracompact \hii\ regions which, according to \teff\ derived from
FIR line ratios, contain stars of spectral types B0-O7 a slight excess
might be found for some cases (Afflerbach et al.\ 1997). However, their
stellar content is not known directly and multiple sources can therefore
not be excluded. The first ultracompact \hii\ region where the spectral
type of the central star has been directly identified (Watson \& Hanson 1997) 
indicates leakage or dust absorption.
Oey (these proceedings) and Oey \& Kennicutt (1997) have recently compared 
observed \hii\ region luminosities with the Lyman continuum luminosities 
predicted for their well determined stellar content. The vast majority
is found to be leaking photons (up to $\sim$ 50 \%).
From their samples, mostly dominated by several O stars, an excess
of Lyman continuum flux can {\em on average} be excluded for O stars.
Similar observations of less populated regions and nebulae dominated by
later spectral types would be highly rewarding.

\section{Photoionization of the diffuse ionized gas by massive stars}
\label{s_pops}
The multi-phase interstellar medium (ISM) contains a diffuse ionized component
representing a significant fraction of its total mass and volume.
This gas is often referred to as the `Reynolds Layer' in the Milky Way, or 
the `Diffuse Ionized Medium' (DIM), `Warm Ionized Medium' (WIM), 
`Diffuse Ionized Gas' (DIG) etc. in external galaxies. 
We shall adopt the acronym `DIG'.
It is usually found that OB stars provide enough power to ionize the DIG
(e.g.\ Reynolds 1984, Hunter \& Gallagher 1990, Ferguson et al. 1996,
Martin \& Kennicutt 1997).
The picture of photoionization by massive stars has, however, been challenged
by recent measurements of a low He ionization fraction in the Galactic DIG
(Heiles et al.\ 1996). On the other hand the observations of Martin \& Kennicutt
(1997) show that important variations in the physical state of the DIG
exist among galaxies.

To provide a quantitative base to address some of the questions raised by
these studies it is highly desirable to use results from the most appropriate
atmosphere and evolutionary models. For example, the expected He ionization
of the DIG is dependent on the hardness of the ionizing fluxes of OB stars, 
which, as discussed above, has been revised considerably in recent years.
Following the lines of Heiles et al.\ (1996) we will therefore derive
new consistent calibrations of the H and \hei\ ionizing fluxes for massive 
star populations based on our {\em CoStar} models.
Some implications for the DIG of various galaxies will be discussed in
Sect.\ \ref{s_dig}

\subsection{Ionizing fluxes of massive star populations}
For studies of the DIG one usually considers a stellar population at a 
steady-state equilibrium (``constant star formation'').
The basic quantities of interest are then (cf.\ Heiles et al.\ 1996):
the lifetime $\tau(m)$ of a star of initial mass $m$,
its time-averaged Lyman continuum luminosity $s_{\rm H}(m)$, and
its time-averaged HeI continuum luminosity $s_{\rm HeI}(m)$.
These quantities can then be integrated over an initial mass function (IMF)
to derive the total output of a stellar population.

For the derivation of these quantities we use the {\em CoStar} models
(papers I-III). We also extend the calculations to include the post 
main-sequence (MS) phases, which have been neglected in all previous
calculations. The adopted input physics (tracks, atmospheres) is the same
as in the synthesis models of Schaerer (1996) and Schaerer \& Vacca (1997):
we use the most recent Geneva tracks, Schmutz et al.\ (1992) 
atmospheres for Wolf-Rayet (WR) stars, and Kurucz (1991) atmospheres for 
the remaining phases.

\begin{figure*}[htb] 
\centerline{
\psfig{width=6.5cm,figure=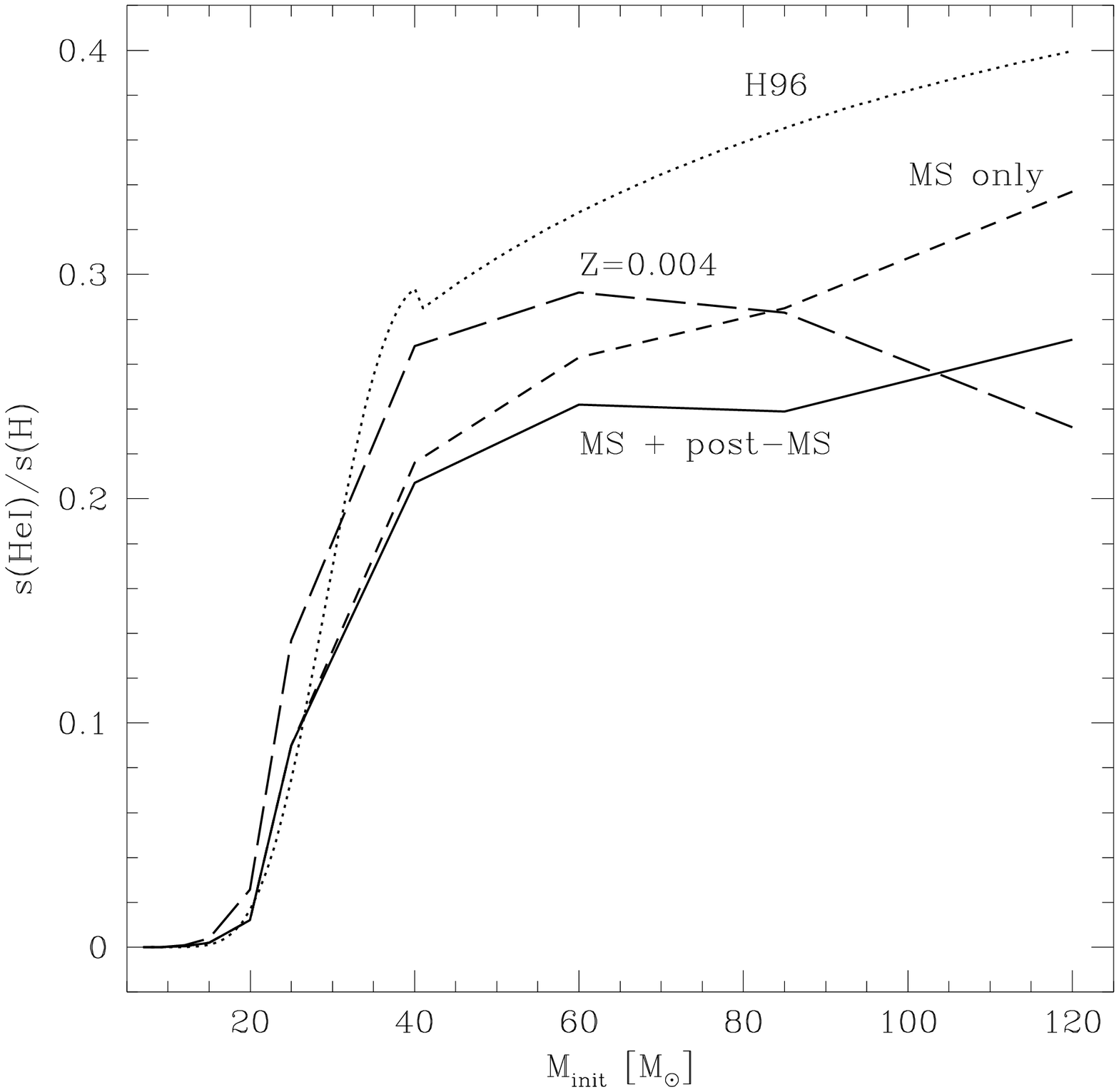}
\psfig{width=6.5cm,figure=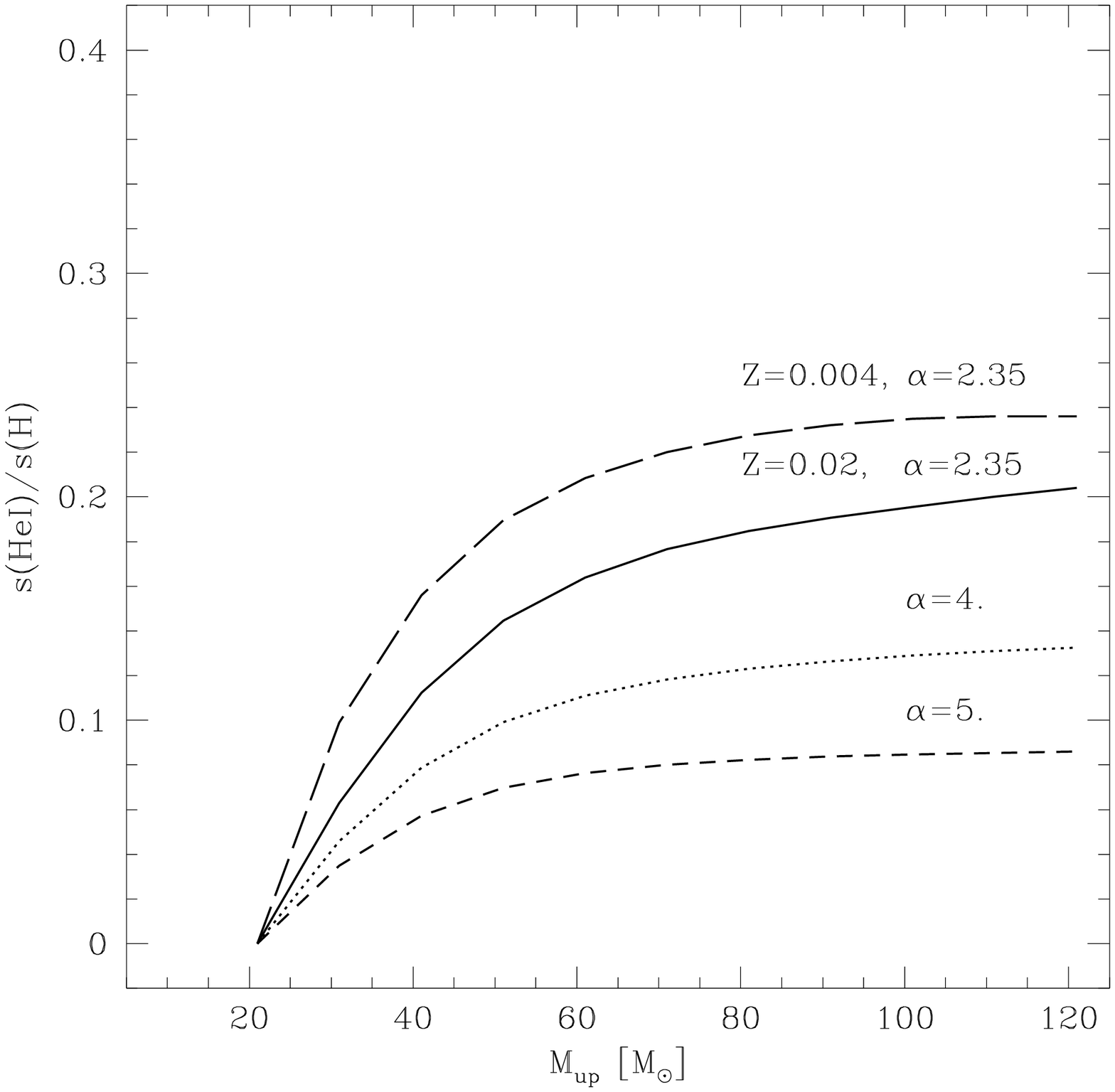}
}
\caption{{\em Left panel:} HeI/H hardness of the ionizing flux as a function of
	initial mass. Time-averaged value over the entire lifetime (solid for solar
	metallicity; long-dashed for 1/5 Z$_\odot$) and over the MS lifetime only 
	(short-dashed for Z$_\odot$) derived from our models. 
	Dotted: Eq.\ 8 of Heiles et al.\ (1996).
{\em Right panel:} HeI/H hardness of the ionizing flux of an integrated steady-state
	stellar population as a function of the upper mass cut-off.
	Solar metallicity models: Salpeter IMF (solid), $\alpha=4.$ (dotted),
	$\alpha=5.$ (short-dashed).
	Model at 1/5 Z$_\odot$ and Salpeter IMF: long-dashed. }
\label{fig_q1q0_mini}
\end{figure*}

Fitting the results for solar metallicity models we obtain the 
following descriptions:
\begin{equation}
	\log \tau_{\rm MS}(m) = 9.818 - 3.286 \log m + 0.8074 (\log m)^2,
\label{eq_tau_ms}
\end{equation}
\begin{equation}
	\log \tau_{\rm total}(m) = 9.986 - 3.496 \log m + 0.8942 (\log m)^2.
\label{eq_tau_tot}
\end{equation}
$\tau_{\rm MS}$ and $ \tau_{\rm total}$ (in years) are the main sequence 
and the total lifetime respectively. 
Eqs.\ \ref{eq_tau_ms} and \ref{eq_tau_tot} are valid for $7 \le m/\msun \le 120$.
Note that the most massive stars 
($m \ge 85 \msun$) enter the WR phase during core H-burning. In this case
we have restricted the definition of the ``MS phase'' to the non-WR phase.
We note that the MS lifetimes differ by less than $\sim$ 15 \% from
the values of G\"usten \& Mezger (1982) used by Heiles et al.
The mean Lyman continuum luminosity can be fitted by
\begin{equation}
	\log s_{\rm H}(m) = 32.47 + 17.35 \log m - 4.362 (\log m)^2,
\label{eq_s_H}
\end{equation}
for the same mass range as above.
It turns out that this expression represents well both the average over the
MS and the total average.
Our values agree well with those of Cox et al.\ (1986) for $m \sim 20-40$ \msun.
Otherwise our values for $s_H$ are lower, with differences up to a factor of $\sim$
5 at the border of the fit interval.
  
To describe the \hei\ ionizing flux $s_{\rm HeI}$ we use 
$\zeta(m) \equiv s_{\rm HeI}/s_{\rm H}$ (see Fig.\ \ref{fig_q1q0_mini}), 
which can be approximated by
\begin{eqnarray}
	\zeta(m) & = &  -0.1604 + 0.0093 \, m  \;\;\;{\mbox {\rm for}} \;\;
20 \le m/M_\odot \le 40 \\
\zeta(m)_{\rm MS} & = &  0.1656 + 0.0014 \,m  \;\;\;{\mbox {\rm for}} \;\; 40 \le m/\msun \le 120 \\
\zeta(m)_{\rm total} & = &  0.1861 + 0.0007 \,m \;\;\;{\mbox {\rm for}} \;\; 40 \le m/\msun \le 120
\end{eqnarray}
For $m < 20$ \msun\ the fraction of \hei\ ionizing photons is negligible 
($\zeta(m) < 0.01$).
%

The ``time-averaged hardness'' $\zeta$ as a function of the initial mass $m$ 
is shown in Fig.\ \ref{fig_q1q0_mini} (left panel). 
Also shown is $\zeta(m)$ given by Heiles et al.\
(1996), which was obtained from the plane parallel non-LTE atmospheres of Kudritzki 
et al.\ (1991). 
Note that these atmosphere models predict a similar hardness (in terms of 
$Q_{\rm HeI}/Q_{\rm H}$) of the ionizing flux as the {\em CoStar} models 
including also wind effects and line blanketing.
The data of Heiles et al., however, only correspond to values typical for characteristic
parameters of main sequence stars; any temporal evolution has been neglected.
The evolutionary tracks in the considered mass range extend to temperatures below 
40000 K on the MS.
Given the rapid decrease of the \hei\ ionizing luminosity below this temperature
(e.g.\ paper III),
the MS average of $\zeta$ is systematically lower than the value given by Heiles 
et al.\ for $m \ga 30 \msun$.
For lower masses our results are in good agreement with the values of Heiles et al.
Incidentally this is also the mass range which is critical for the interpretation
of the observations of the DIG in the Milky Way (see Sect.\ \ref{s_dig}).
It must, however, also be noted that this is likely the domain where
current models are the most uncertain for the reasons discussed in paper III.
Fig.\ \ref{fig_q1q0_mini} also illustrates the difference between averaging
over the MS only and the more correct average over the MS and post-MS phases:
For the most massive stars $\zeta$ averaged over their entire lifetime
can be up to $\sim$ 25 \% lower than the MS value. This is due to their evolution
in the WR phase, which leads to lower average luminosities.
For comparison we have also plotted $\zeta$ (average over total lifetime) at 
1/5 solar metallicity. As expected the average spectrum is slightly harder, mostly
due to the temperature shift of the MS.
{\em 
In short we conclude that the time-averaged spectrum predicted from consistent 
modeling using in particular the ionizing fluxes from our {\em CoStar} models 
yields a lower HeI/H hardness than the results of Heiles et al.\ 
(1996)}.\footnote{This contradicts several statements of Martin \& Kennicutt (1997).
Some comments must be made on the mass scales shown in their Fig.\ 3:
1) All scales are representative for dwarf stars and do not include
the time average, whose importance was shown above.
2) Scale d was drawn from the preprint version of Heiles et al.\ which used
LTE model atmospheres, while their final results are based on 
non-LTE models yielding a similar hardness as the {\em CoStar} models.
Similarly for other results ($M_{\rm up}$, Galactic star formation rate) 
the statements of Martin \& Kennicutt refer to the preprint of Heiles et al.,
before important revisions were made.}

Let us now consider the hardness $\zeta$ of the spectrum produced by an
integrated stellar population at a steady-state equilibrium.
Figure \ref{fig_q1q0_mini} illustrates the dependence of $\zeta$ on the 
upper mass cut-off $M_{\rm up}$ for various slopes $\alpha$ of the IMF.
In our notation the Salpeter IMF is $\alpha=2.35$. 
Obviously the results are not sensitive to the lower mass cut-off 
$M_{\rm low}$ as long as $M_{\rm low} \la 20 \msun$.
Despite differences in all the ingredients (see above) our results 
agree well with those of Heiles et al.\ (their Table 6)
for $M_{\rm up} \la 30 \msun$, whereas we obtain a softer spectrum for 
larger values of $M_{\rm up}$.
For illustration we also plot the run of $\zeta$ for a low metallicity 
population.
We shall now briefly discuss some implications from these results.

\subsection{Implications for the diffuse ionized gas}
\label{s_dig}
As shown by Heiles et al.\ (1996) the observations of the DIG in
the Milky Way pose serious difficulties for the massive-star photoionization
picture:
while the high fraction of neutral He requires that very massive stars 
are excluded, the observed Galactic ionization requirement would
imply a total star formation rate much larger than derived from other
observations.
Indeed the ratio of the HeI ionizing photon production rate $Q({\rm HeI})$ 
to $Q({\rm H})$ derived from the observations is $\la 0.013-0.027$ 
(Heiles et al.). 

{\em Our results confirm entirely the interpretation of their Milky Way data.}
In contrast with some statements of Martin \& Kennicutt the use of
our atmosphere models does not change the conclusions of Heiles et al.\ 
for the reasons discussed above (see footnote). Indeed,
a comparison with the right panel of Fig.\ \ref{fig_q1q0_mini} shows
that the observed $Q({\rm HeI})/Q({\rm H})$ implies an upper mass cut-off 
of $M_{\rm up} \sim 25 \msun$. 
This result is robust even to large variations of the IMF slope 
(e.g.\ the very steep slopes found by Massey et al.\ (1995) for field 
stars) as also show by Fig.\ \ref{fig_q1q0_mini}.
For $\alpha=2.35$ we also obtain the same Lyman continuum production as 
Heiles et al., which confirms the requirement of a large Galactic
star formation rate to reproduce the observed Galactic ionizing photon
production rate.

As an unconventional explanation for these problems Heiles et al.\
consider the possible excess of Lyman continuum flux from B stars,
motivated by the observed excess in the B2II star $\epsilon$ CMa.
If common among B stars this could potentially solve the above mentioned
difficulties. As discussed in Sect.\ \ref{s_lyc} it should be possible to 
constrain the ionizing fluxes of B stars observationally in the 
future.

The observations of DIG of three Magellanic irregular galaxies of 
Martin \& Kennicutt (1997) reveal a different picture. 
There the He ionization fraction is approximately equal to the hydrogen
fraction, which translates to a lower limit of $Q({\rm HeI})/Q({\rm H})
\ga 0.1$ (Martin \& Kennicutt). From the right panel in Fig.\ 
\ref{fig_q1q0_mini} this translates to a lower limit for the upper
mass cut-off of $M_{\rm up} \ga$ 30-40 \msun\ for a Salpeter IMF.
We also note that the high He ionization in these objects could in 
principle even be explained by a population with an IMF slope
as steep as those derived by Massey et al.\ (1995) for field stars.
For this case the required upper mass cut-off is $M_{\rm up} \ga$ 40-60
\msun, well compatible with numerous independent observational constraints 
(cf.\ Leitherer 1997, Massey 1997).
The ionization by radiation leaking out from \hii\ regions (with
a ``normal'' IMF slope) is, however, a more likely explanation, 
as shown by recent observations (Ferguson et al.\ 1996, Martin \& 
Kennicutt 1997, Oey \& Kennicutt 1997, Wang et al.\ 1997).

\section{Reliable ionizing fluxes for OB stars: further developments}
\label{s_new}
To improve our atmosphere models further and to compare the predicted 
spectra with UV observations we have recently upgraded the non-LTE 
calculations by including a detailed treatment of CNO and Si in addition 
to H and He (following de Koter et al.\ 1997).
Improvements are also brought about for the completeness of the line list
used in the line blanketing Monte-Carlo simulations.
In this Section we shall briefly present some preliminary results
on the EUV fluxes and compare our latest calculations with
new models from the Munich group (Pauldrach, these proceedings). 

\begin{figure*}[htb]
\centerline{
\psfig{width=6.5cm,figure=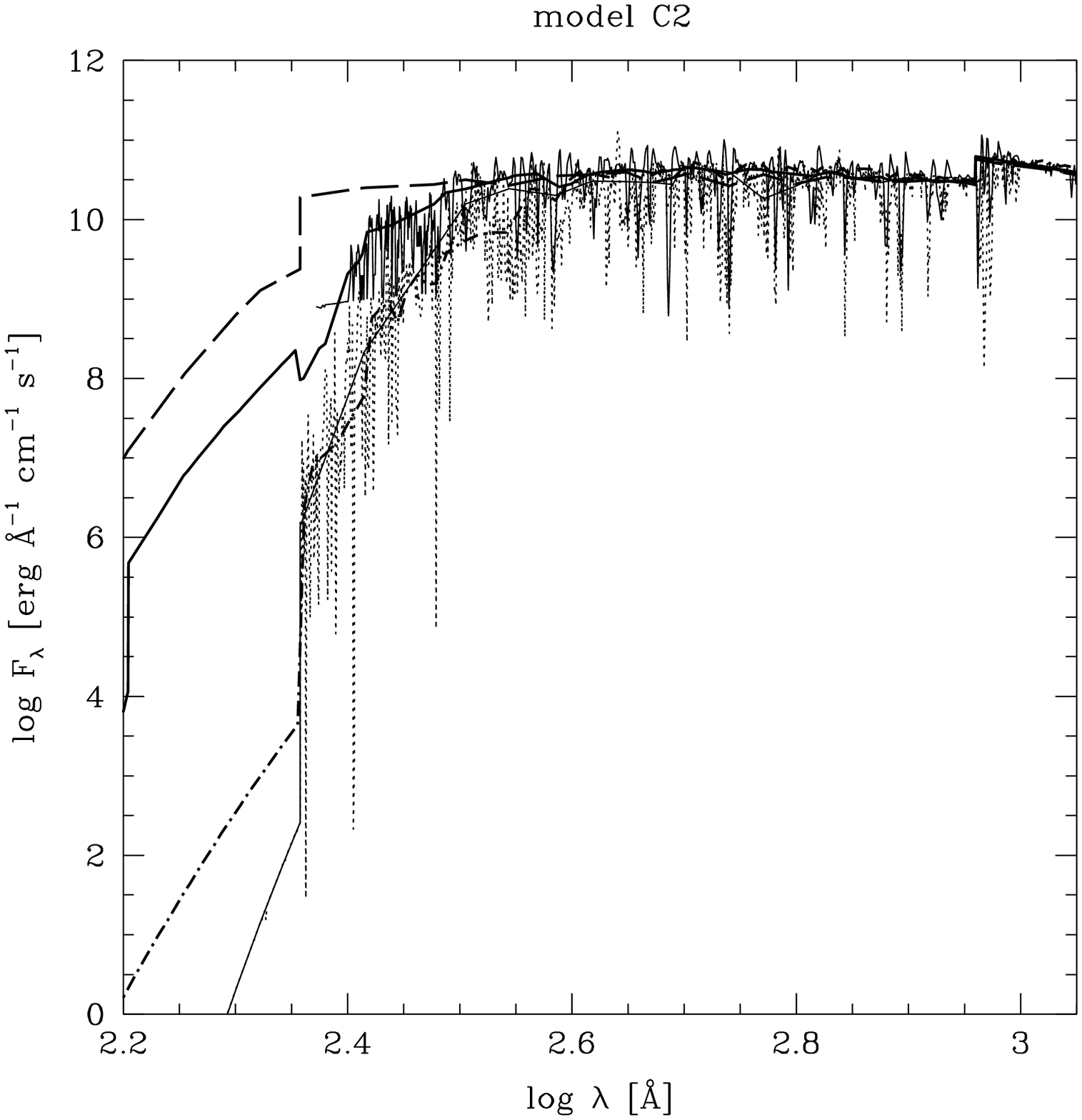}
\psfig{width=6.5cm,figure=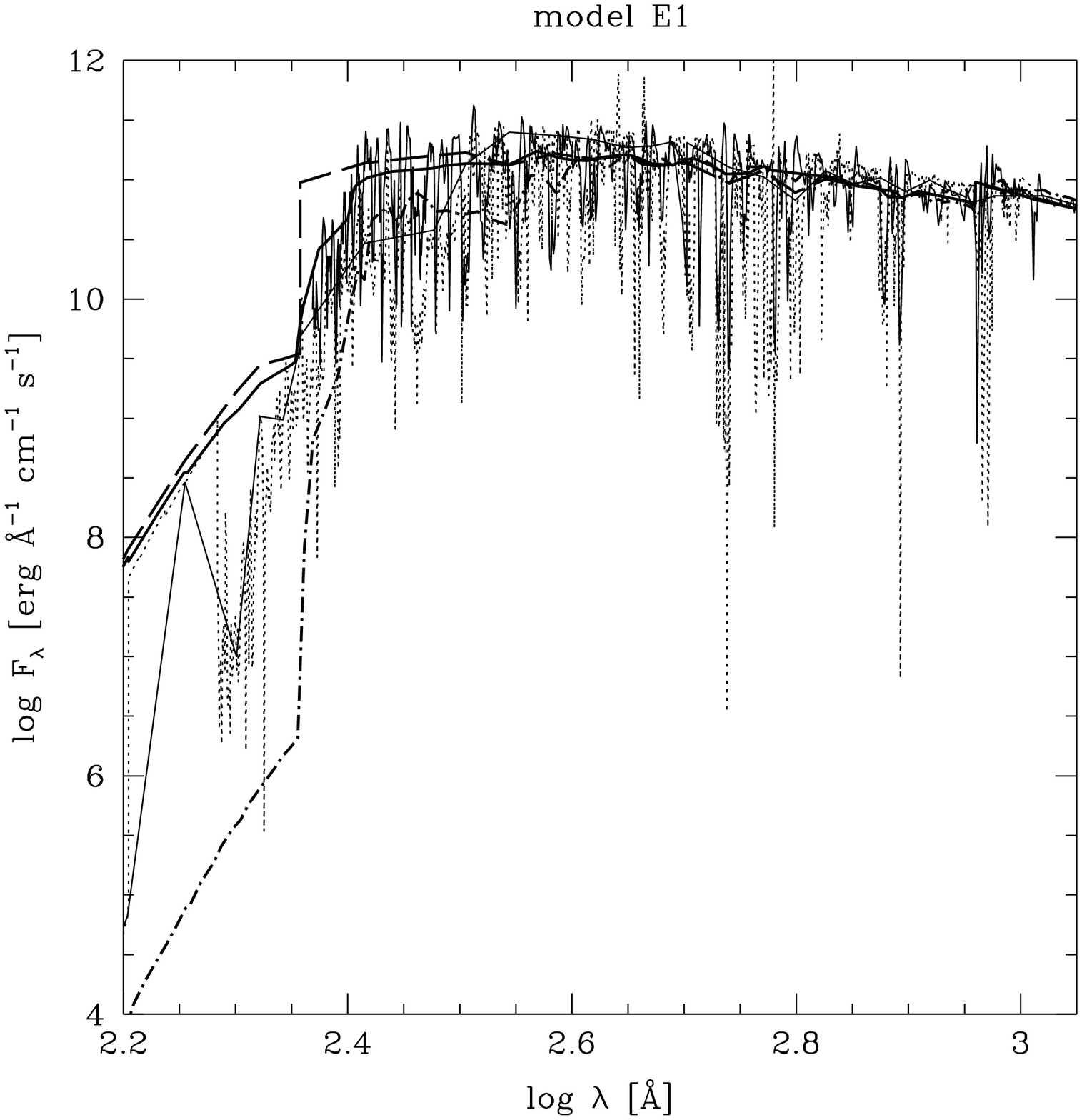}
}
\caption{Comparison of the emergent fluxes from different models.
Long-dashed: from paper IV, Solid: new model (continuous distribution: thick,
detailed line+cont.: thin), Dotted: Pauldrach model (smoothed spectrum 
traced by thin solid line), Dashed-dotted: Kurucz model.
{\em Left panel:} model C2 (\teff\ $\sim$ 40000 K),
{\em right panel:} model E1 (\teff\ $\sim$ 50000 K),
}
\label{fig_new}
\end{figure*}


The main changes with respect to paper IV are the explicit non-LTE
treatment of CNO and Si, whose populations were previously described
by our modified nebular approximation (see paper I and references therein).
By the same token this also allows us to derive a finer description
of the radiation temperature as a function of the ionization potential,
which is used to describe the ionization structure of the remaining
elements. The bound-free opacities of CNO and Si are now also included.
In Fig.\ \ref{fig_new} we plot a comparison for two models from paper IV.
For $\lambda \ga 310$ \AA\ the new calculations agree well with the previous
ones. Between $\sim$ 310 \AA\ and the \heii\ edge (228 \AA) our new 
models predict less flux, mostly due to the inclusion of the bound-free 
CNO opacity. 
Predictions for the \heii\ continuum remain subject to other considerable
uncertainties (see paper IV).

In Fig.\ \ref{fig_new} we have also plotted the flux from two new 
atmosphere models including line blocking calculated with similar parameters 
(see Pauldrach, these proceedings).
The overall agreement between the models is quite good for $\lambda \ga 310$
\AA, although some distinct features exist. At shorter wavelengths, however,
the flux from the Munich model decreases more strongly, remaining fairly close
to the Kurucz model for \teff\ $\sim$ 40000 K. (left panel).
We have done test calculations of photoionization models using both sets of 
fluxes. Regarding the nebular Ne$^{++}$/O$^{++}$ abundance ratio (cf.\ Sect.\
\ref{s_hii}) we obtained only minor changes for our models, while the Pauldrach 
model at 40000 K yield Ne$^{++}$/O$^{++}$ $\sim$ 0.1-0.15 and 
O$^{++}$/S$^{++}$ $\ga$ 25, more similar to Kurucz models.

Given important differences between the model techniques and the some of
the input physics (e.g.\ line lists) it is difficult to explain the reason(s)
for the differences in the predicted spectra.
More detailed work remains to be done to further explore the dependence
on various model assumptions and to improve the treatment of line blanketing
and non-LTE effects in spherically expanding atmospheres.
Concomitantly it appears very useful to pursue further studies of ``simple''
\hii\ regions to constrain atmosphere models as much as possible.
In view of the difficulties hot star models have faced so far in reproducing
correctly the observable UV from consistent calculations (see e.g.\
Groenewegen \& Lamers 1991, Pauldrach et al.\ 1994) such new tests could be 
crucial to reach the goal of obtaining accurate predictions 
of the spectral evolution of early type stars.

{\small
\acknowledgements{I would like to thank Alex de Koter for the use of
his ISA-WIND code, Grazyna Stasi\'nska for model calculations, and Adi
Pauldrach for sharing results from new calculations.
A fellowship from the Swiss National 
Foundation of Scientific Research and partial support from the 
Directors Discretionary Research Fund of the STScI are greatly
acknowledged.}
}
\end{document}